\documentclass[twocolumn,showpacs,preprintnumbers,APSl,prd,nofootinbib,superscriptaddress]{revtex4-1}
\usepackage{dcolumn}
\usepackage{bm}
\usepackage{ifpdf}
\usepackage{hyperref}
\usepackage{dcolumn}
\usepackage{bm}
\usepackage[spanish,english]{babel}
\usepackage{amsfonts}
\usepackage{amssymb}
\usepackage{graphicx}
\usepackage[latin1]{inputenc}

\begin{document}

\title{Topological vortices in generalized Born-Infeld-Higgs electrodynamics}
\author{R. Casana}
\email{rodolfo.casana@gmail.com}
\affiliation{Departamento de F\'isica, Universidade Federal do Maranh\~ao, 65080-805,
S\~ao Lu\'is, Maranh\~ao, Brazil.}
\author{E. da Hora}
\affiliation{Departamento de F\'isica, Universidade Federal do Maranh\~ao, 65080-805,
S\~ao Lu\'is, Maranh\~ao, Brazil.}
\affiliation{Coordenadoria Interdisciplinar de Ci\^encia e Tecnologia, Universidade
Federal do Maranh\~ao, 65080-805, S\~ao Lu\'is, Maranh\~ao, Brazil.}
\author{D. Rubiera-Garcia}
\email{drubiera@fudan.edu.cn}
\affiliation{Center for Field Theory and Particle Physics and Department of Physics,
Fudan University, 220 Handan Road, 200433 Shanghai, China}
\author{C. dos Santos}
\affiliation{Centro de Física e Departamento de Física e Astronomia, Faculdade de Ciê%
ncias da Universidade do Porto, 4169-007 Porto, Portugal.}
\date{\today }

\begin{abstract}
A consistent BPS formalism to study the existence of topological axially
symmetric vortices in generalized versions of the Born-Infeld-Higgs
electrodynamics is implemented. Such a generalization modifies the field
dynamics via introduction of three non-negative functions depending only in
the Higgs field, namely, $G(|\phi|)$, $w(|\phi|) $ and $V(|\phi|)$. A set of
first-order differential equations is attained when these functions satisfy
a constraint related to the Ampere law. Such a constraint allows to minimize
the system energy in such way that it becomes proportional to the magnetic
flux. Our results provides an enhancement of topological vortex solutions in
Born-Infeld-Higgs electrodynamics. Finally, we analyze a set of models such
that a generalized version of Maxwell-Higgs electrodynamics is recovered in
a certain limit of the theory.
\end{abstract}

\maketitle

\section{Introduction}

The well-known Born-Infeld electrodynamics was originally introduced to
remove the divergence of electron's self-energy in classical electrodynamics
by introducing a square-root form of the Lagrangian density replacing the
standard Maxwell Lagrangian \cite{BI}. In this way the field strength tensor
remains bounded everywhere and the energy associated to a point-like charge
becomes finite. This theory is a distinguished member of the family of
nonlinear electrodynamics since it enjoys three properties: i) Maxwell
weak-field limit, ii) electric-magnetic duality \cite{em-duality}, and iii)
absence of shock waves and birefringence phenomena concerning propagation of
waves, belonging to the class of theories called ``completely exceptional"
\cite{wave}. Applications of Born-Infeld electrodynamics within gravitation
and cosmology have been considered for many years \cite{BI-applications}.
This model is moreover deemed of a special attention since it appears in the
low-energy limit of string/D-Branes physics \cite%
{D-branes1,D-branes2,D-branes3}.

On the other hand, the study of magnetic vortices gained great interest
since Abrikosov's description for Type-II superconductors \cite{Abrikosov},
which arise naturally from the non-relativistic limit of Ginzburg-Landau
(GL) theory \cite{GL}. In field theory, stable vortex configurations came up
with the seminal work by Nielsen and Olesen \cite{NO} whose study of the
Maxwell-Higgs (MH) model shows that electrically neutral vortex solutions
correspond to the ones obtained by Abrikosov. Lately it was verified the
existence of electrically charged vortex solutions in Chern-Simons-Higgs
(CSH) \cite{CS,CSV} and Maxwell-Chern-Simons-Higgs (MCSH) \cite{MCS,Bolog}
models. In all these cases, the presence of the Higgs fields is essential
for the existence of vortex-type solutions.

Recently, it has been intensively studied the existence of topological
defects in generalized or new effective field theories. For example one can
introduce noncanonical kinetic terms \cite{o1,Hora_Rubiera}, in order to
circumvent the constraints of Derrick's theorem \cite{Derrick} and obtain
topological defect solutions (see e.g. \cite{Books} for a more detailed
account on soliton-like solutions in field theory). Other models are defined
by introducing generalizing functions on standard field models \cite{o2}. In
some cases these generalized models provide self-dual analytical solutions
which certainly enriches our understanding of the field \cite{PLB,AHEP}.
Moreover, this procedure allows to control properties of the topological
defect, such as its width or energy density, providing valuable models for
the analysis of several physical problems. In the literature there are many
interesting applications of these new solutions within several different
scenarios, specially involving the accelerated inflationary phase of the
universe \cite{n8} via the so-called k-essence models \cite{shutt}, strong
gravitational waves \cite{sgw}, tachyon matter \cite{tm}, dark matter \cite%
{dm}, and other topics \cite{o}.

Among generalized models the simplest ones are those generalizing the
Maxwell-Higgs model \cite{gmh}, Chern-Simons-Higgs model \cite{gcsh} and
Maxwell-Chern-Simons-Higgs model \cite{gmcsh}. Based on earlier work on
vortices in Born-Infeld-Higgs models \cite{Shiraishi}, in Ref. \cite%
{Hora_Rubiera} a generalization of Born-Infeld-Maxwell-Higgs (BIMH) model
was constructed within the context of generalized dynamics, but self-dual or
BPS vortices were not found. The main aim of the present manuscript is to
show the existence of self-dual topological BPS vortices in a generalized
BIMH electrodynamics, and study their properties.

\section{Generalized Born-Infeld vortices}

The Lagrangian density of our ($2+1$)-dimensional theory is written as
\begin{equation}
\mathcal{L}=\beta ^{2}\left( 1-\mathcal{R}\right) +w\left( \left\vert \phi
\right\vert \right) \left\vert D_{\mu }\phi \right\vert ^{2}-W\left(
\left\vert \phi \right\vert \right) \text{,}  \label{eq:action}
\end{equation}%
with the definitions
\begin{eqnarray}
\mathcal{R} &=&\sqrt{1+\frac{G\left( \left\vert \phi \right\vert \right) }{%
2\beta ^{2}}F_{\mu \nu }F^{\mu \nu }}, \\
W\left( \left\vert \phi \right\vert \right) &=&\beta ^{2}\left[ 1-V\left(
\left\vert \phi \right\vert \right) \right] ,
\end{eqnarray}%
where $F_{\mu \nu }=\partial _{\mu }A_{\nu }-\partial _{\nu }A_{\mu }$ is
the field strength tensor of the vector potential $A_{\mu }$, while the
covariant derivative realizing the coupling between the gauge and Higgs
fields is given by $D_{\mu }\phi =\partial _{\mu }\phi -ieA_{\mu }\phi $.
The positive functions $G\left( \left\vert \phi \right\vert \right) $ and $%
w\left( \left\vert \phi \right\vert \right) $ are the generalizing functions
in the kinetic sector. The generalized potential $W\left(
\left\vert \phi \right\vert \right) $, a nonnegative function, inherits its structure from the function $V(|\phi |)$, which is restricted by the
condition $0<V(|\phi |)\leq 1$, so $W(\phi)>0$. The Born-Infeld parameter, $\beta $%
, provides modified dynamics for both scalar and gauge fields further
enriching the family of possible models.

From the action (\ref{eq:action}) the gauge field equations of motion read%
\begin{equation}  \label{eq:gauge}
\partial _{\nu }\left( \frac{G}{\mathcal{R}}F^{\nu \mu }\right) =ewJ^{\mu },
\end{equation}%
where $J^{\mu }=i(\phi \partial ^{\mu }\phi ^{\ast }-\phi ^{\ast }\partial
^{\mu }\phi) -2eA^{\mu }\left\vert \phi \right\vert ^{2}$ plays the role of
a current.

At static regime, Eq. (\ref{eq:gauge}) provides the Gauss law,
\begin{equation}
\partial _{j}\left( \frac{G}{\mathcal{R}}\partial _{j}A_{0}\right)
=2we^{2}A_{0}\left\vert \phi \right\vert ^{2},  \label{eq:Gauss}
\end{equation}%
which is saturated by temporal gauge: $A_{0}=0$. In this way we see that the
model, at static regime and in temporal gauge, describes electrically
neutral magnetic configurations. Under these conditions, from Eq. (\ref%
{eq:gauge}) the Ampere law reads
\begin{equation}
\epsilon _{kj}\partial _{j}\left( \frac{G}{\mathcal{R}}B\right) -ewJ_{k}=0,
\label{eq:Ampere}
\end{equation}%
and Higgs's field equation becomes
\begin{eqnarray}
0 &=&w\left( D_{j}D_{j}\phi \right) +\left( \partial _{j}w\right) D_{j}\phi
\nonumber \\[-0.1cm]
&&  \label{eq:Higgs} \\[-0.1cm]
&&-\frac{\partial w}{\partial \phi ^{\ast }}\left\vert D_{j}\phi \right\vert
^{2}-\frac{B^{2}}{\ 2\mathcal{R}}\frac{\partial G}{\partial \phi ^{\ast }}-%
\frac{\partial W}{\partial \phi ^{\ast }},  \nonumber
\end{eqnarray}%
where in the last two equations $\mathcal{R}$ reads
\begin{equation}
\mathcal{R}=\left( 1+\frac{G}{\beta ^{2}}B^{2}\right) ^{1/2}.  \label{RR}
\end{equation}

The energy-momentum tensor of the model is given by
\begin{eqnarray}
T_{\mu \nu } &=&-\frac{1}{\mathcal{R}}G\left( \left\vert \phi \right\vert
\right) F^{\beta }{}_{\mu }F_{\beta \nu }+w\left( \left\vert \phi
\right\vert \right) \left( D_{\mu }\phi \right) ^{\ast }D_{\nu }\phi
\nonumber \\[0.2cm]
&&+w\left( \left\vert \phi \right\vert \right) \left( D_{\nu }\phi \right)
^{\ast }D_{\mu }\phi -g_{\mu \nu }\mathcal{L}.  \label{tem}
\end{eqnarray}

In this work we are interested in searching for electrically neutral
magnetic vortices and, more specifically, we will study such solutions at
static regime and in temporal gauge. As it is largely known in literature,
the axially symmetric vortex ansatz works fine to find such solutions,
namely,
\begin{equation}
\phi =vg\left( r\right) e^{in\theta },~A_{\theta }=-\frac{a\left( r\right) -n%
}{er},  \label{eq:ansatz}
\end{equation}%
where $n$ is an integer number and $a(r)$ and $g(r)$ are regular functions
that satisfy the following boundary conditions
\begin{eqnarray}
g\left( 0\right) &=&0~,~\ \ g\left( \infty \right) =1  \label{eq:bound1} \\%
[0.15cm]
a\left( 0\right) &=&n~,~\ \ a\left( \infty \right) =0  \label{eq:bound2}
\end{eqnarray}%
Using this ansatz the magnetic field is written as%
\begin{equation}
B=-\frac{a^{\prime }}{er}.  \label{BB}
\end{equation}%
with the short-hand notation $a^{\prime }\equiv da/dr$.

For the ansatz (\ref{eq:ansatz}) the Ampere law (\ref{eq:Ampere}) is
expressed as
\begin{equation}
\left( \frac{G}{\mathcal{R}}B\right) ^{\prime }=-2ev^{2}w\frac{ag^{2}}{r},
\label{ampere}
\end{equation}%
while Higgs' field equation (\ref{eq:Higgs})\ reads
\begin{eqnarray}
0 &=&g^{\prime \prime }+\frac{g^{\prime }}{r}-\frac{a^{2}}{r^{2}}g+\frac{1}{%
2w}\frac{dw}{dg}\left[ \left( g^{\prime }\right) ^{2}-\left( \frac{ag}{r}%
\right) ^{2}\right]  \nonumber  \label{eq:higgs3} \\
&&  \label{higgs} \\
&&-\frac{1}{4wv^{2}}\frac{B^{2}}{\mathcal{R}}\frac{dG}{dg}-\frac{1}{2wv^{2}}%
\frac{dW}{dg},  \nonumber
\end{eqnarray}%
where $\mathcal{R}$ is given by Eq.(\ref{RR}).

\subsection{The BPS formalism}

The energy of the vortex is given by the integration of the $T_{00}$
component of (\ref{tem}) which, in static regime and in the gauge $%
A_{0}=0$, is given by

\begin{equation}
T_{00}=\beta ^{2}\left( \mathcal{R}-V\right) +wv^{2}\left( g^{\prime
}\right) ^{2}+wv^{2}\left( \frac{ag}{r}\right) ^{2},
\end{equation}%
and will be nonnegative whenever the condition $\mathcal{R}\geq V$ is
satisfied. The total energy reads
\begin{equation}
E=2\pi \!\int \!\!drr\!\left[ \beta ^{2}\left( \mathcal{R}-V\right)
+wv^{2}\left( g^{\prime }\right) ^{2}+wv^{2}\left( \frac{ag}{r}\right) ^{2}%
\right] ,  \label{E1}
\end{equation}%
where the fields were expressed in terms of the ansatz (\ref{eq:ansatz}). We
now use the Bogomol'nyi trick \cite{Bogol} to rewrite it as
\begin{eqnarray}
E &=&2\pi \!\int \!\!drr\left[ \frac{1}{2}\frac{G}{\mathcal{R}}\left( B\mp
\beta \sqrt{\frac{2F}{G}}\right) ^{2}+wv^{2}\left( g^{\prime }\mp \frac{ag}{r%
}\right) ^{2}\right.  \nonumber \\
&&\hspace{2cm}\pm B\frac{\beta }{\mathcal{R}}\sqrt{2FG}\pm 2wv^{2}\frac{ag}{r%
}g^{\prime }  \label{E2} \\
&&\hspace{2cm}\left. +\beta ^{2}\mathcal{R}-\frac{1}{2}\frac{GB^{2}}{%
\mathcal{R}}-\frac{\beta ^{2}F}{\mathcal{R}}-\beta ^{2}V\right] ,  \nonumber
\end{eqnarray}%
where we have introduced the function $F$, which is, in principle, arbitrary
but nonnegative, to be determined later in order to obtain
solutions with well defined energy. Using the definition (\ref{RR}) in the
third row, we can rewrite (\ref{E2}) as
\begin{eqnarray}
E &=&2\pi \!\int \!\!drr\left\{ \pm B\frac{\beta }{\mathcal{R}}\sqrt{2FG}\pm
2wv^{2}\frac{ag}{r}g^{\prime }\right.  \label{xqa} \\[0.2cm]
&&\hspace{0cm}+\frac{1}{2}\frac{G}{\mathcal{R}}\left( B\mp \beta \sqrt{\frac{%
2F}{G}}\right) ^{2}+wv^{2}\left( g^{\prime }\mp \frac{ag}{r}\right) ^{2}
\nonumber \\[0.2cm]
&&\hspace{0cm}\left. +\frac{\beta ^{2}}{\mathcal{R}}\left[ \frac{1}{2}\left(
\mathcal{R}-V\right) ^{2}+\frac{1}{2}\left( 1-V^{2}\right) -F\right]
\right\} .  \nonumber
\end{eqnarray}%
We observe that, by imposing the expression in the third row to be null, this
allows to determine the function $F$ in terms of $V$ and
$R$, namely
\begin{equation}
F=\frac{1}{2}\left( \mathcal{R}-V\right) ^{2}+\frac{1}{2}\left(
1-V^{2}\right) ,  \label{XX}
\end{equation}%
which shows that $F$ is a nonnegative function because $0<V\leq 1$. Let us point out that the function $F$ is defined without
considering the self-dual equations or the BPS limit.

Now by considering condition (\ref{XX}) and the expression (\ref{BB}%
) for the magnetic field, the energy (\ref{xqa}) reads%
\begin{eqnarray}
E &=&2\pi \!v^{2}\int \!\!drr\left\{ \mp \frac{a^{\prime }}{r}\left( \frac{%
\beta \sqrt{2FG}}{ev^{2}\mathcal{R}}\right) \mp \frac{a}{r}\left(
-2wgg^{\prime }\right) \right.  \label{xqb} \\
&&\hspace{0.5cm}\left. +\frac{1}{2}\frac{G}{\mathcal{R}}\left( B\mp \beta
\sqrt{\frac{2F}{G}}\right) ^{2}+wv^{2}\left( g^{\prime }\mp \frac{ag}{r}%
\right) ^{2}\right\} .  \nonumber
\end{eqnarray}

We can now transform the first term\ in a total derivative by setting
\begin{equation}
\left( \frac{\beta }{ev^{2}\mathcal{R}}\sqrt{2FG}\right) ^{\prime
}=-2wgg^{\prime },  \label{cd1}
\end{equation}%
This way, the energy (\ref{xqb})\ is written as
\begin{eqnarray}
E &=&2\pi \!v^{2}\!\int \!\!drr\left[ \mp \frac{1}{r}\left( a\Omega \right)
^{\prime }\right. \\
&&~\ \ \ \ \ ~\ \ \ \left. \hspace{-1cm}wv^{2}\left( g^{\prime }\mp \frac{ag%
}{r}\right) ^{2}+\frac{1}{2}\frac{G}{\mathcal{R}}\left( B\mp \beta \sqrt{%
\frac{2F}{G}}\right) ^{2}\right] ,  \nonumber
\end{eqnarray}%
where
\begin{equation}
\Omega =\frac{\beta }{ev^{2}\mathcal{R}}\sqrt{2FG},
\end{equation}%
with the function $F$ given by Eq. (\ref{XX}). We can
now further constrain the set of functions $G$, $w$
and $V$ in order to attain a true lower bound for the energy by
selecting functions satisfying
\begin{equation} \label{eq:constraint}
\Omega \left( 0\right) =1,~\Omega \left( \infty \right) =\text{finite}
\label{oo}
\end{equation}%
Then, by considering the boundary conditions given by Eqs. (\ref{oo}%
) and (\ref{eq:bound1}), the energy reads
\begin{eqnarray}
E &=&2\pi v^{2}\left\vert n\right\vert  \label{qxqx} \\
&&\hspace{-1cm}+2\pi \!\int \!\!drr\left[ wv^{2}\left( g^{\prime }\mp \frac{%
ag}{r}\right) ^{2}+\frac{1}{2}\frac{G}{\mathcal{R}}\left( B\mp \beta \sqrt{%
\frac{2F}{G}}\right) ^{2}\right] .  \nonumber
\end{eqnarray}%
This clearly shows that the energy possess a lower bound
\begin{equation} \label{eq:bound}
E\geq 2\pi v^{2}\left\vert n\right\vert ,
\end{equation}%
whenever the functions $G$, $w$ and $V$ chosen provide a function $\Omega $ satisfying Eq. (\ref{oo}). Such a
lower bound is saturated when the fields satisfy the BPS\ or self-dual
equations
\begin{equation}
g^{\prime }=\pm \frac{ag}{r},  \label{BPS2_0}
\end{equation}%
\begin{equation}
B=\pm \frac{\beta }{V}\sqrt{\frac{1-V^{2}}{G}}.  \label{BPS2_1}
\end{equation}%
This is a set of first-order equations that satisfy automatically the
second-order equations (\ref{ampere}) and (\ref{eq:higgs3}), as can be
immediately seen by derivation of the former. This is so because the
Euler-Lagrangian equations only imply that a static BPS solution will be a
stationary point of the energy. In Eq.(\ref{BPS2_1}) we have used (\ref{XX})
to compute $F$ in the BPS limit, which provides
\begin{equation}
F=\frac{1-V^{2}}{2V^{2}},
\end{equation}%
a nonnegative function due to $0<V\leq 1$. Similarly, the
nonnegative function $\Omega \left( r\right) $ is given by
\begin{equation}
\Omega _{bps}=\beta \sqrt{G\left( 1-V^{2}\right) }.
\end{equation}

By using the BPS equations (\ref{BPS2_0}) and (\ref{BPS2_1}), Ampere's law (%
\ref{ampere}) can be written as
\begin{equation}
\frac{d}{dg}\sqrt{\beta ^{2}G\left( 1-V^{2}\right) }=-2ev^{2}wg.
\label{xcxc}
\end{equation}%
This relation allows to determine one of the generalizing functions when the
other two are given, for example, we can compute $w$ if we provide the
functions $G$ and $V$. Here it is worthwhile to notice that Eq.(\ref{xcxc})
is exactly the condition (\ref{cd1}) in the BPS\ limit.

To conclude this section, the BPS energy density of the model, which appears
in
\begin{equation}
E_{bps}=\displaystyle{2\pi \!\int \!\!drr~\varepsilon _{bps},}
\end{equation}%
is given by
\begin{equation}
\varepsilon _{bps}=\frac{\beta ^{2}}{V}\left( 1-V^{2}\right) +2wv^{2}\left(
\frac{ag}{r}\right) ^{2}.
\end{equation}%
and will be positive definite whenever the functions $0<V\left( g\right)
\leq 1$ and $w\left( g\right) \geq 0$\textbf{.}

\section{A family of models}

In this section we shall focus on the special case
\begin{equation}
V\left( \left\vert \phi \right\vert \right) =\sqrt{1-\frac{2U\left(
\left\vert \phi \right\vert \right) }{\beta ^{2}}},
\end{equation}%
since this choice, in the limit $\beta \rightarrow \infty $, allows to
obtain the generalized Maxwell-Higgs model from the Lagrangian density (\ref%
{eq:action}):
\begin{equation}
\mathcal{L}=-\frac{G\left( \left\vert \phi \right\vert \right) }{4}F_{\mu
\nu }F^{\mu \nu }+w\left( \left\vert \phi \right\vert \right) \left\vert
D_{\mu }\phi \right\vert ^{2}-U\left( \left\vert \phi \right\vert \right) .
\end{equation}

With this choice the BPS\ equations read
\begin{eqnarray}
g^{\prime } &=&\pm \frac{ag}{r},~  \label{BPS_1} \\
B &=&\pm \sqrt{\frac{2U}{G}}\left( 1-\frac{2U}{\beta ^{2}}\right) ^{-1}.
\label{BPS_2}
\end{eqnarray}
The condition (\ref{xcxc}) reads
\begin{equation}
\frac{d}{dg}\sqrt{2UG}=-2ev^{2}wg,  \label{cxcx}
\end{equation}%
and the BPS energy density is
\begin{equation}
\varepsilon _{bps}=2U\left( 1-\frac{2U}{\beta ^{2}}\right)
^{-1/2}+2v^{2}w\left( \frac{ag}{r}\right) ^{2}.
\end{equation}%
Therefore, the generalized models can be defined by choosing $G\left(
g\right) $ and $U\left( g\right) $ functions which, via the constraint (\ref%
{cxcx}), allow to find the remaining function $w\left( g\right) $. These
three functions must be nonnegative for positive definiteness of the energy
density. In the next sections we shall choose some models satisfying the constraint (\ref{eq:constraint}) and, therefore, their BPS solutions will saturate the bound (\ref{eq:bound}).

\subsection{Some choices for the potential}

Next we shall consider two classes of models characterized by the form of
the \textquotedblleft potential" $~U(g)$. First we will consider, in each
case, the asymptotic behavior of the functions $g(r)$ and $a(r)$ compatible
with the boundary conditions that make the energy finite and positive, and
next solve the BPS equations. On the other hand we note that $\beta$ is not
a constant characterizing the solutions but rather a parameter determining a
particular model within the family defined by the corresponding term in the
action (\ref{eq:action}). In some of the following numerical cases we shall
treat nevertheless $\beta$ as a free parameter for the computations, which
means that in those cases we will be comparing the behavior of the solutions
corresponding to different models of the family of generalized Born-Infeld
Lagrangians. In order to perform the numerical analysis, without loss of
generality, we set $e=1=v$.

\subsection{Asymptotic behavior for $\left\vert \protect\phi \right\vert
^{4} $ models}

The $\left\vert \phi \right\vert ^{4}$-models are described by the function $%
U\left( g\right) $\ given by
\begin{equation}
U\left( g\right) =\frac{1}{2}\left( 1-g^{2}\right) ^{2},
\end{equation}%
and the function $G\left( g\right) $ whose behavior when $r\rightarrow 0$ is
\begin{equation}
G\left( g\right) =\alpha _{0}+\alpha _{2}g^{2}+...,
\end{equation}%
and when $r\rightarrow \infty $ reads%
\begin{equation}
G\left( g\right) =\alpha _{0}^{\left( \infty \right) }+\alpha _{1}^{\left(
\infty \right) }\left( 1-g\right) +\alpha _{2}^{\left( \infty \right)
}\left( 1-g\right) ^{2}+....
\end{equation}
where $\alpha_0, \alpha_2, \ldots$, and $\alpha _{0}^{\left( \infty \right)
}, \alpha _{1}^{\left( \infty \right) }, \ldots$ are some constants.

By introducing the above information into the BPS equations (\ref{BPS_1})
and (\ref{BPS_2}), we can compute the behavior of field profiles when $%
r\rightarrow 0$:
\begin{eqnarray}
g\left( r\right) &=&C_{n}r^{n}-\left( \frac{{\beta }^{2}}{{\beta }^{2}-1}%
\right) ^{1/2}\frac{C_{n}r^{n+2}}{4\left( \alpha _{0}\right) ^{1/2}}+... \\
a\left( r\right) &=&n-\left( \frac{{\beta }^{2}}{{\beta }^{2}-1}\right)
^{1/2}\frac{r^{2}}{2\left( \alpha _{0}\right) ^{1/2}} \\
&&\hspace{-1cm}+\left( \frac{{\beta }^{2}}{{\beta }^{2}-1}\right) ^{3/2}%
\frac{\left( 2\alpha _{0}+\alpha _{2}\right) {\beta }^{2}-\alpha _{2}}{%
\left( \alpha _{0}\right) ^{3/2}{\beta }^{2}}\frac{{C}_{n}^{2}{r}^{2n+2}}{%
4\left( n+1\right) }+...  \nonumber
\end{eqnarray}
where $C_n$ is a set of constants.

Similarly, we calculate the behavior of the profiles at infinity:
\begin{eqnarray}
1-g\left( r\right) &\sim &r^{-1/2}\exp \left[ -r\sqrt{2}\left( \alpha
_{0}^{\left( \infty \right) }\right) ^{-1/4}\right] \\
a\left( r\right) &\sim &r^{1/2}\exp \left[ -r\sqrt{2}\left( \alpha
_{0}^{\left( \infty \right) }\right) ^{-1/4}\right].
\end{eqnarray}
These expansions are fully consistent with the assumed boundary conditions (%
\ref{eq:bound1}) and (\ref{eq:bound2}) for the BPS solutions.

\subsection{Asymptotic behavior for $\left\vert \protect\phi \right\vert
^{6} $ models}

In this case, the function $U\left( g\right) $ is given by
\begin{equation}
U\left( g\right) =\frac{1}{2}g^{2}\left( 1-g^{2}\right) ^{2}.
\end{equation}

We consider the behavior of function $G$, which at origin takes the form
\begin{equation}
G\left( g\right) =\frac{\gamma _{-2}}{g^{2}}+\gamma _{0}+\gamma
_{2}g^{2}+...,
\end{equation}%
and at infinity is%
\begin{equation}
G\left( g\right) =\gamma _{0}^{\left( \infty \right) }+\gamma _{1}^{\left(
\infty \right) }\left( 1-g\right) +\gamma _{2}^{\left( \infty \right)
}\left( 1-g\right) ^{2}+...
\end{equation}

The behavior of the profiles at $r\rightarrow 0$ is%
\begin{eqnarray}
g\left( r\right) &\sim &C_{n}r^{n}-\frac{C_{n}^{3}r^{3n+2}}{4\left( \gamma
_{-2}\right) ^{1/2}\left( n+1\right) ^{2}}+... \\
a\left( r\right) &\sim &n-\frac{{C}_{n}^{2}r^{2n+2}}{2\left( \gamma
_{-2}\right) ^{1/2}\left( n+1\right) } \\
&&+\frac{\left( 2\gamma _{-2}+\gamma _{0}\right) \beta ^{2}-\gamma _{-2}}{%
4\beta ^{2}\left( \gamma _{-2}\right) ^{3/2}\left( 2n+1\right) }{C}_{n}^{4}{r%
}^{4\,n+2}+...  \nonumber
\end{eqnarray}

while the asymptotic behavior for $r\rightarrow \infty $ is
\begin{eqnarray}
1-g\left( r\right) &\sim &r^{-1/2}\exp \left[ -r\sqrt{2}\left( \gamma
_{0}^{\left( \infty \right) }\right) ^{-1/4}\right] \\
a\left( r\right) &\sim &r^{1/2}\exp \left[ -r\sqrt{2}\left( \gamma
_{0}^{\left( \infty \right) }\right) ^{-1/4}\right]
\end{eqnarray}
and, again, these expansions are consistent with the problem under
consideration.

\subsection{Discussion of results}

\begin{figure}[t]
\centering\includegraphics[width=8.5cm]{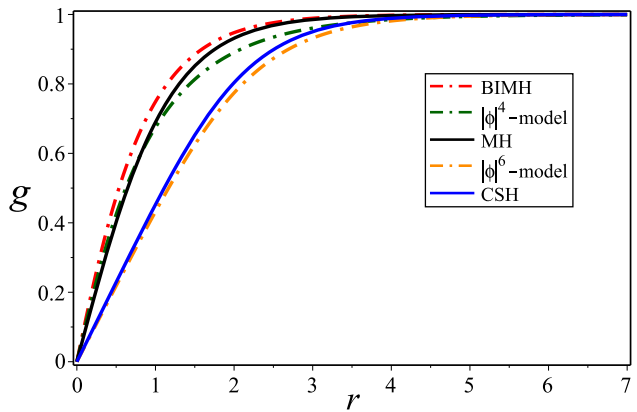}
\caption{ Higgs field ${g}(r)$, for $n=1$ and dashed-dotted lines for $%
\protect\beta =1.5$. }
\label{higgs_p}
\end{figure}

\begin{figure}[t]
\centering\includegraphics[width=8.5cm]{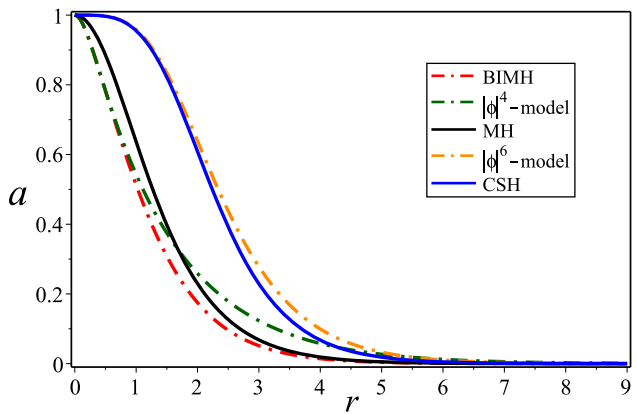}
\caption{Gauge field profile ${a}(r)$, for $n=1$ and dashed-dotted lines for
$\protect\beta =1.5$.}
\label{gauge_p}
\end{figure}

\begin{figure}[t]
\centering\includegraphics[width=8.5cm]{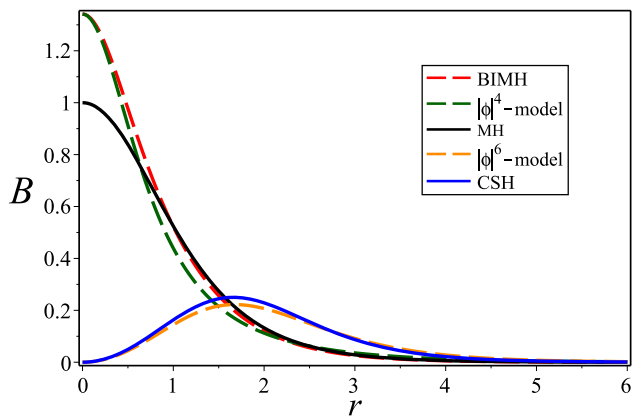}
\caption{Magnetic field profiles ${B}(r)$, for $n=1$ and dashed lines for $%
\protect\beta=1.5$.}
\label{magnetic_p}
\end{figure}

\begin{figure}[t]
\centering\includegraphics[width=8.5cm]{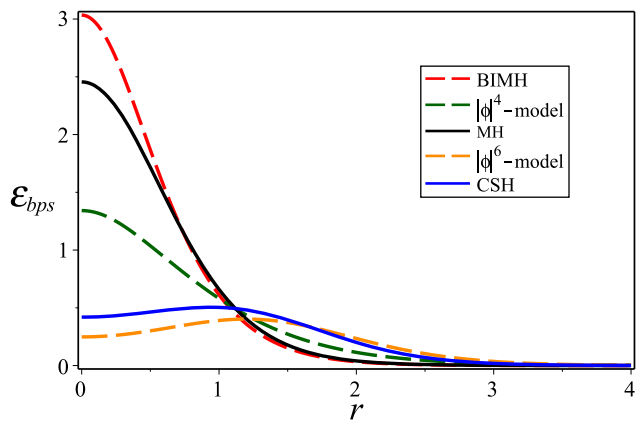}
\caption{BPS energy density ${\protect\varepsilon}_{_{bps}}(r)$, for $n=1$
and dashed lines for $\protect\beta=1.5$.}
\label{energy_p}
\end{figure}

\begin{figure}[t]
\includegraphics[width=8.5cm]{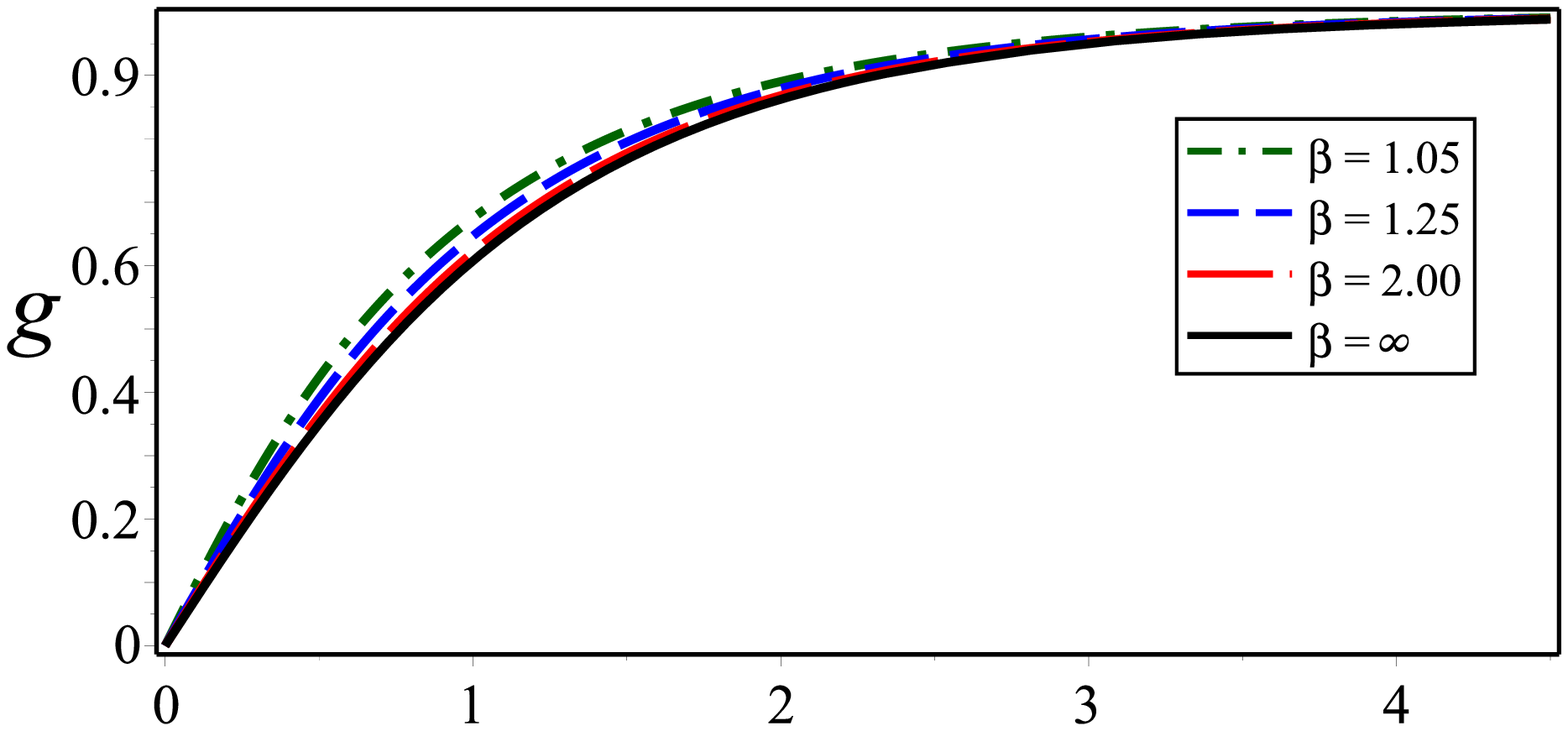} %
\includegraphics[width=8.5cm]{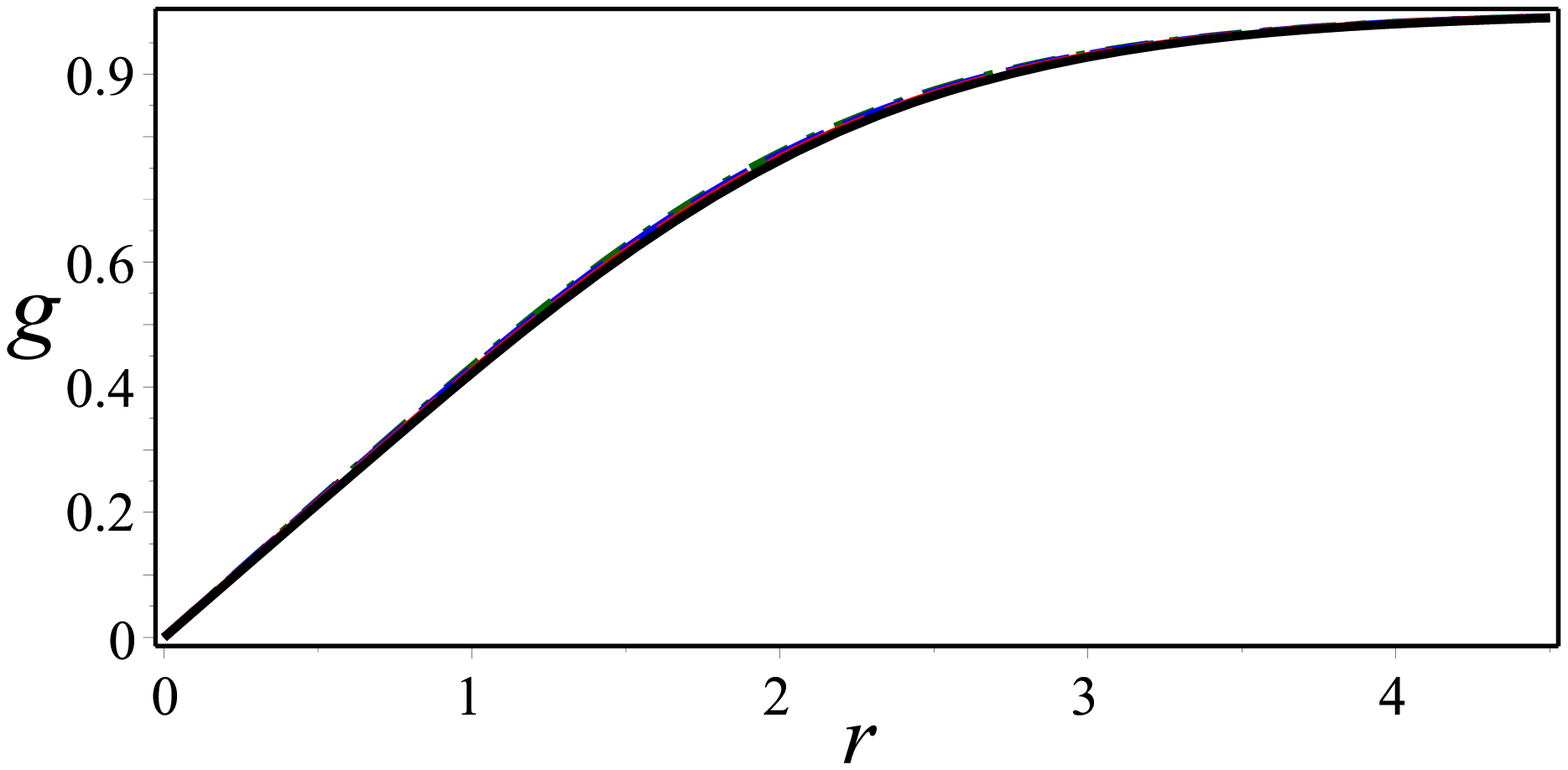}
\caption{Higgs field profile $g(r)$, for $n=1$. Upper figure represents the $%
|\protect\phi|^4$-model defined by Eq. (\protect\ref{eq:model2}) and the
bottom figure represents the $|\protect\phi|^6$-model defined in Eq. (%
\protect\ref{eq:model3}).}
\label{higgs_2_3}
\end{figure}

\begin{figure}[t]
\hspace{0.1cm} \includegraphics[height=5.5cm,
width=4.0cm]{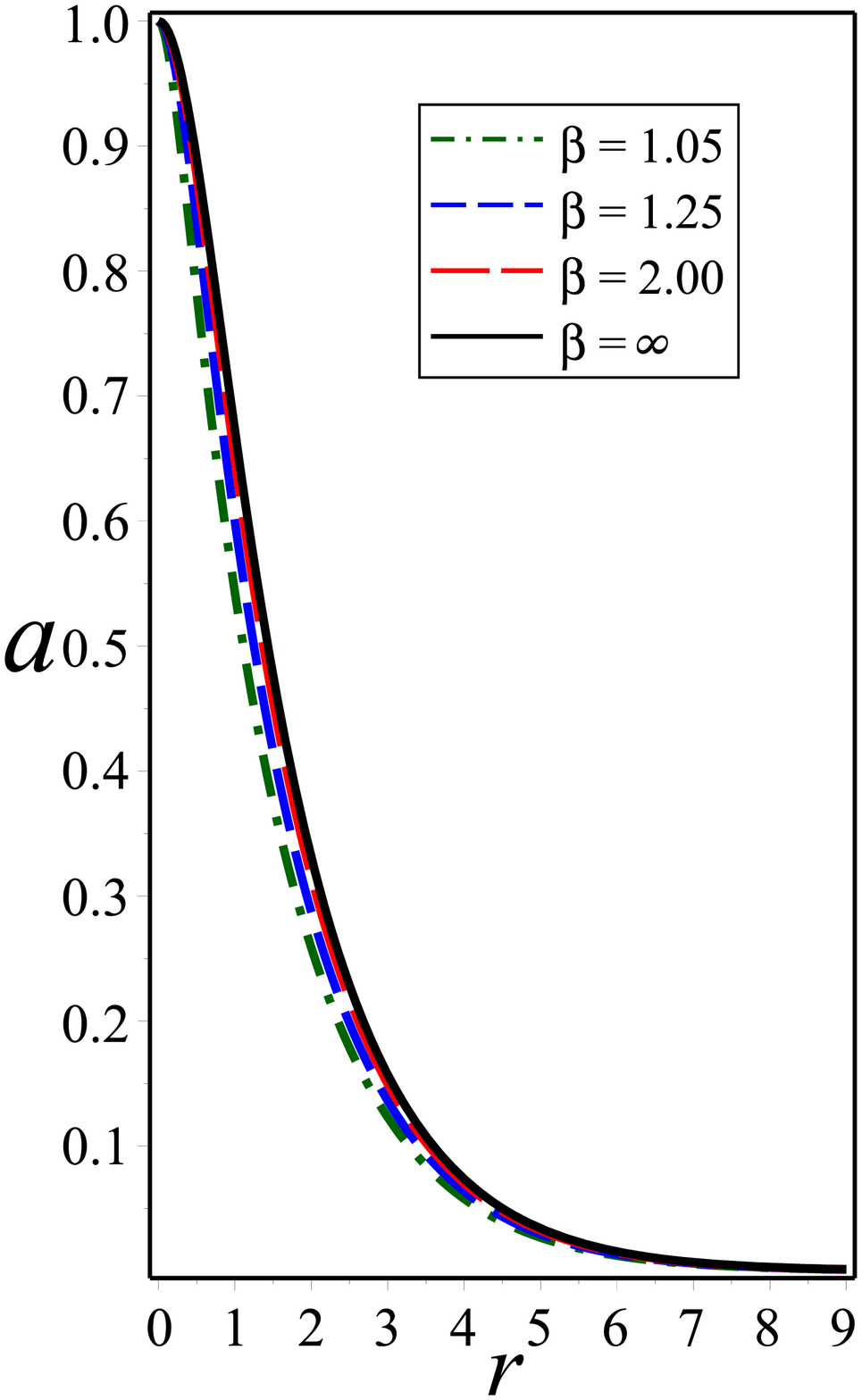}\includegraphics[height=5.5cm,
width=4.0cm]{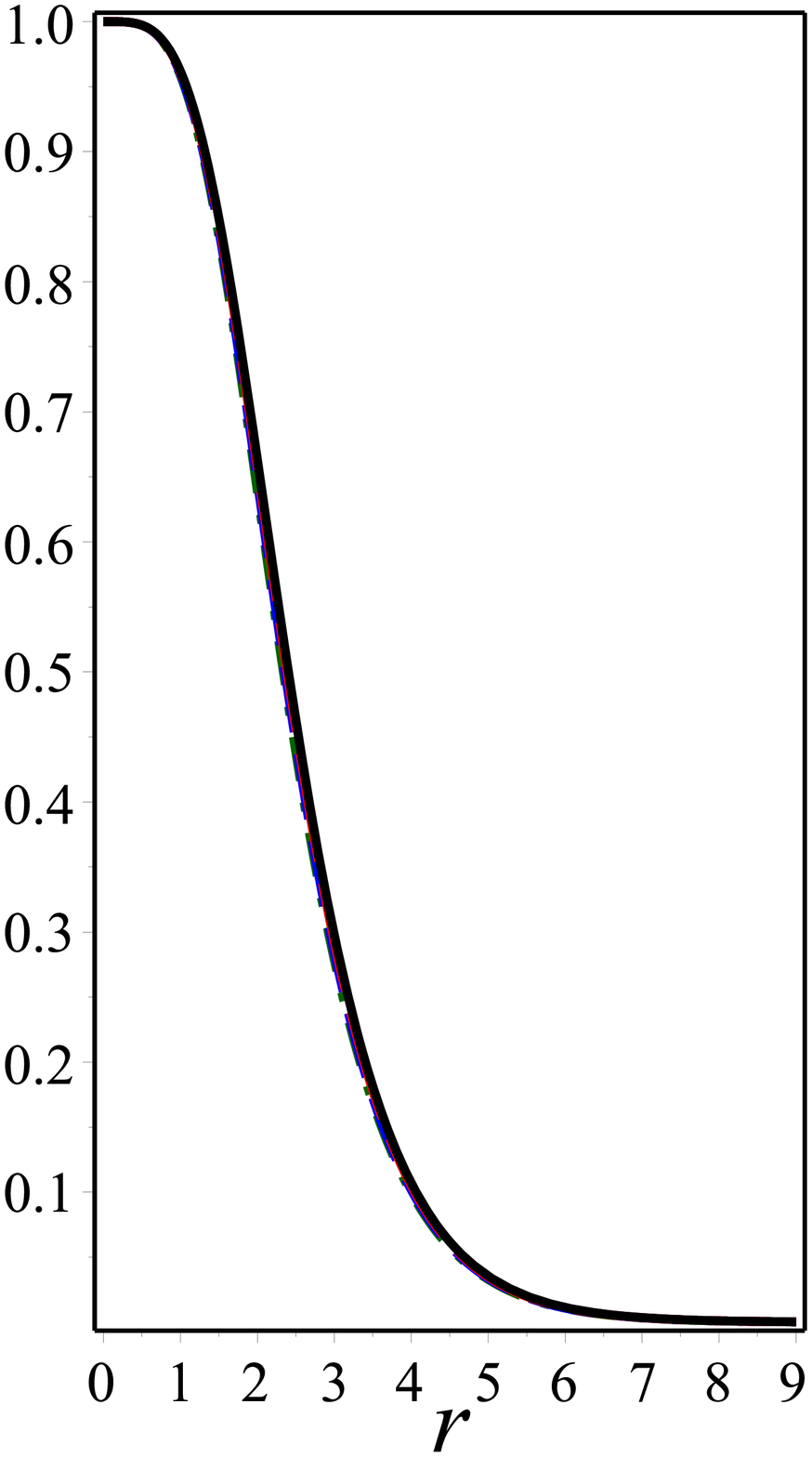}
\caption{Gauge field profile $a(r)$, for $n=1$. Left figure represents the $|%
\protect\phi|^4$-model defined by Eq. (\protect\ref{eq:model2}) and
right-figure represents the $|\protect\phi|^6$-model defined in Eq. (\protect
\ref{eq:model3}).}
\label{vetorial_2_3}
\end{figure}

\begin{figure}[t]
\hspace{0.1cm} \includegraphics[height=5.5cm,
width=4.0cm]{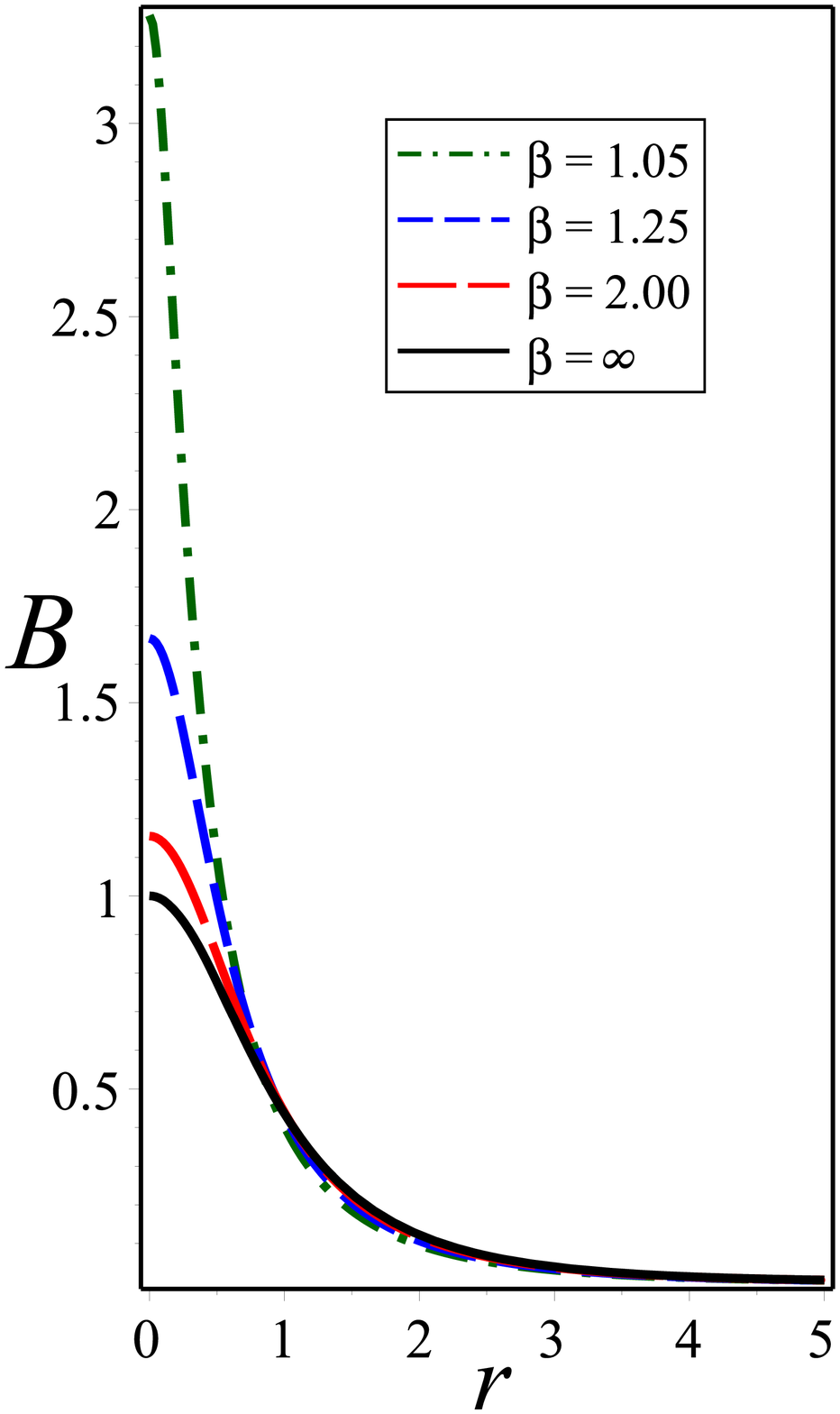}\includegraphics[height=5.5cm,
width=4.0cm]{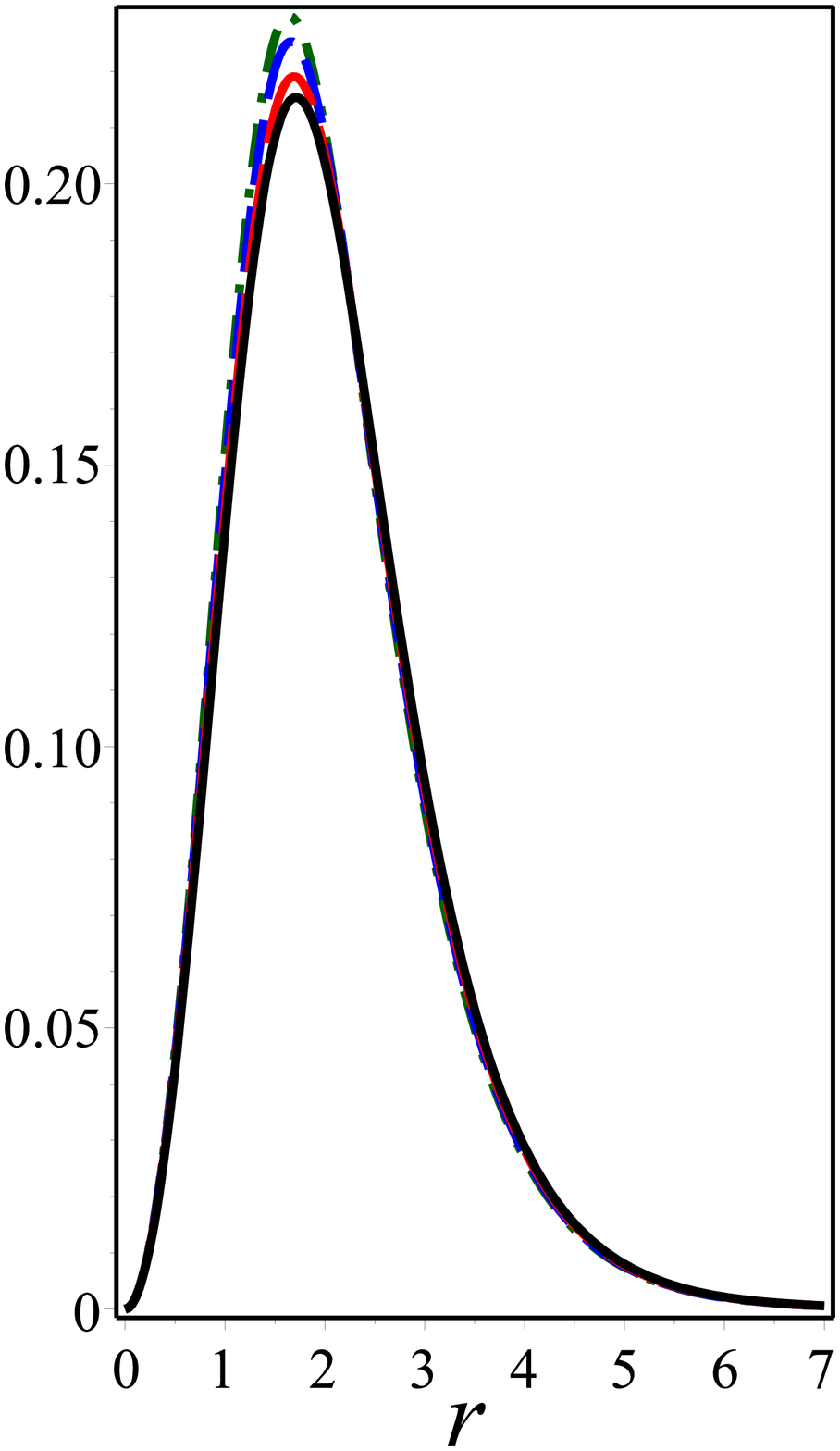}
\caption{Magnetic field $B(r)$, for $n=1$. Left figure represents the $|%
\protect\phi|^4$-model defined by Eq. (\protect\ref{eq:model2}) and right
figure represents the $|\protect\phi|^6$-model defined in Eq. (\protect\ref%
{eq:model3}).}
\label{magnetic_2_3}
\end{figure}

\begin{figure}[t]
\includegraphics[height=5.5cm, width=4.0cm]{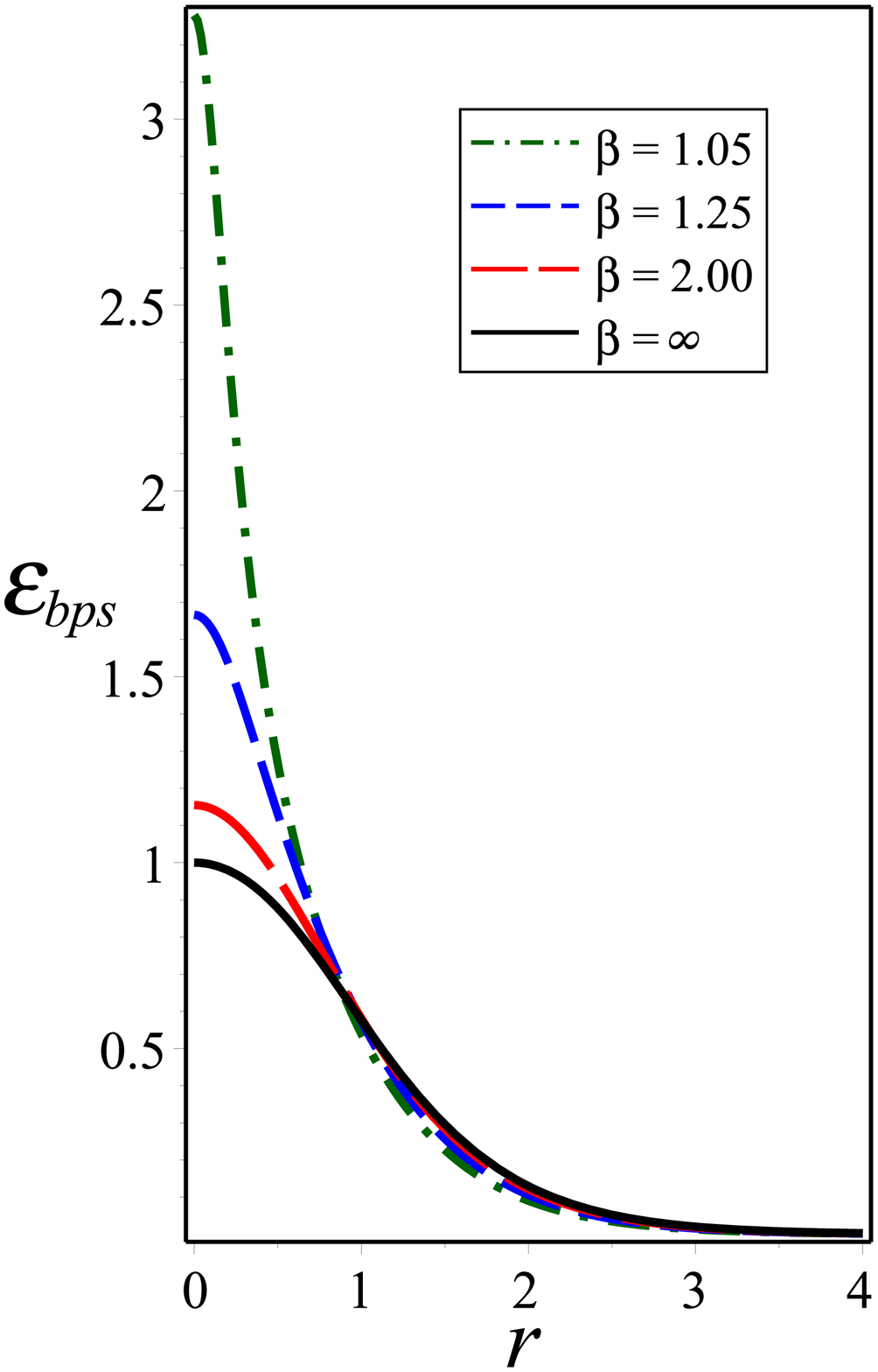}%
\includegraphics[height=5.5cm, width=4.0cm]{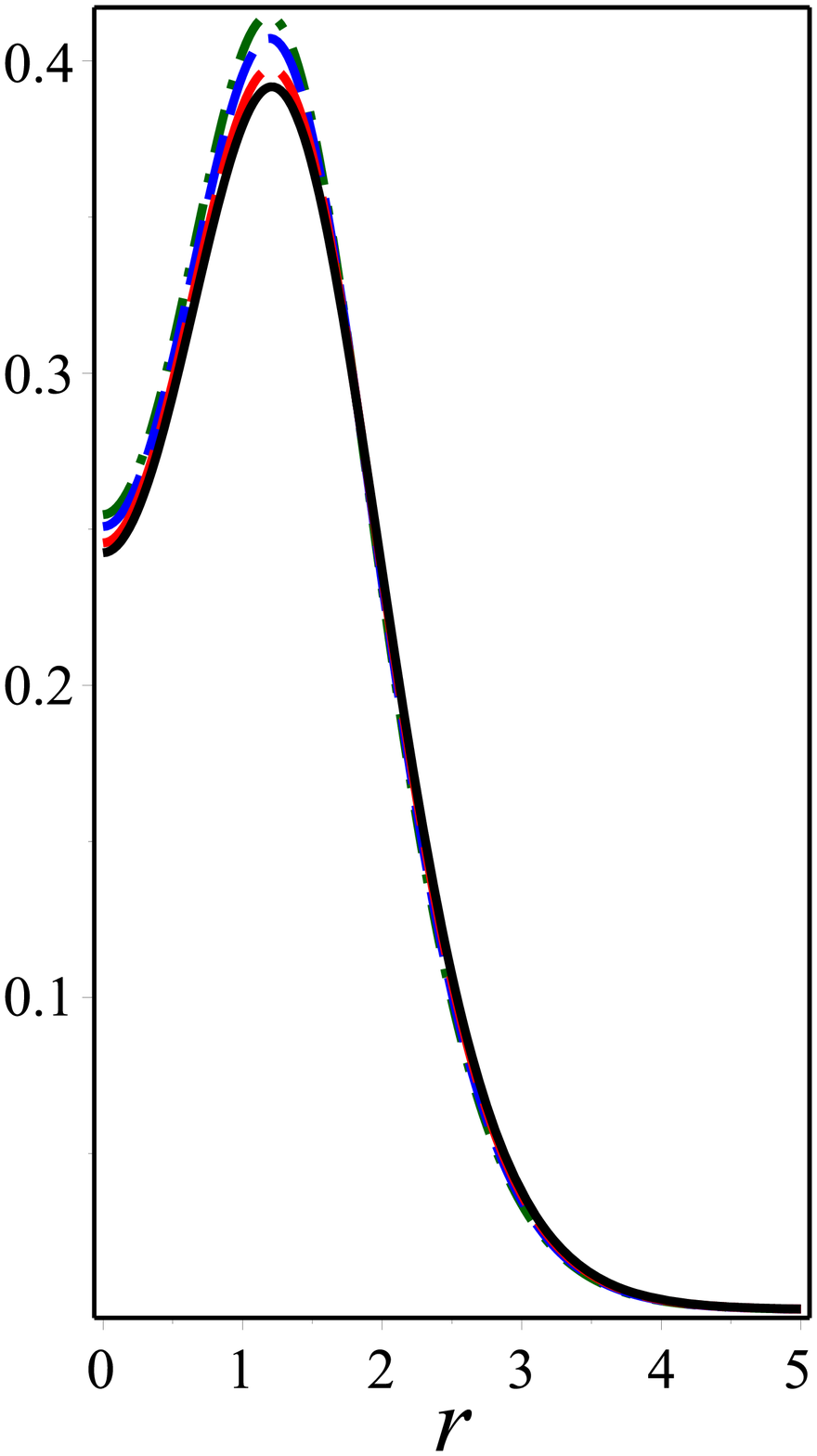}
\caption{BPS energy density $\protect\varepsilon_{bps}(r)$, for $n=1$. Left
figure represents the $|\protect\phi|^4$-model defined by Eq. (\protect\ref%
{eq:model2}) and right figure represents the $|\protect\phi|^6$-model
defined in Eq. (\protect\ref{eq:model3}).}
\label{energy_2_3}
\end{figure}

Once the boundary conditions are fixed, we have performed numerical
solutions of the BPS equations (\ref{BPS_1}) and (\ref{BPS_2}) by using
routines of Maple 16.2. The first numerical results are obtained by
considering fixed values of $\beta(=1.05)$, and comparing the standard MH,
CSH and BIMH models with our $\left\vert \phi \right\vert ^{4}$-BIMH and $%
\left\vert \phi \right\vert^{6}$-BIMH models. These results are shown in
Figs. \ref{higgs_p}, \ref{gauge_p}, \ref{magnetic_p} and \ref{energy_p}. The
second numerical analysis was performed by fixing $n=1$ and varying the
values of $\beta(=1.05, 1.25, 2.00, \infty$), with the resulting profiles
depicted in Figs. \ref{higgs_2_3}, \ref{vetorial_2_3}, \ref{magnetic_2_3}
and \ref{energy_2_3} for the $\left\vert \phi \right\vert^{4}$-BIMH and $%
\left\vert \phi \right\vert^{6}$-BIMH models studied in this work. In both
scenarios, we have depicted the field profiles $g(r)$, $a(r)$, the magnetic
field $B(r)$ and the BPS energy density $\varepsilon _{bps}(r)$
corresponding to the different models under comparison.


To further clarify the plots, we note that the first $\phi ^{4}$-model is
defined by
\begin{equation}
G=1,~w=1,  \label{eq:model1}
\end{equation}
and represents the standard Born-Infeld-Maxwell-Higgs model (dashed-dotted
red lines in Figs. \ref{higgs_p}--\ref{energy_p}).

The second one (dashed-dotted green lines in Figs. \ref{higgs_p}--\ref%
{energy_p}) is given by the following functions%
\begin{equation}
G\left( g\right) =\exp \left( 2g^{2}\right) ,\ w\left( g\right) =g^{2}\exp
\left( 2g^{2}\right)  \label{eq:model2}
\end{equation}%
The $\phi ^{6}$-model (dashed-dotted orange lines in Figs. \ref{higgs_p}--%
\ref{energy_p}) is defined by the functions
\begin{equation}
G\left( g\right) =\frac{\left( 3+g^{2}\right) ^{2}}{9g^{2}},~w\left(
g\right) =\frac{2}{3}\left( 1+g^{2}\right) .  \label{eq:model3}
\end{equation}
For completeness, we also depict the profiles of the standard MH (solid
black line) and CSH models (solid blue line).

In general we see that the introduction of a finite value for $\beta $ has a
non-trivial impact on the profiles of $a(r)$ and $g(r)$. This follows from
the comparison between the standard $\phi ^{4}$ BIMH model in Eq. (\ref%
{eq:model1}) (red dashed curve, corresponding to $\beta =1.5$) and the
standard MH system (solid black curve), with the former vortex being ticker
than the latter. We also see that the impact of changing the $G$ and $\omega
$ functions through the new $\left\vert \phi \right\vert^{4}$ and $%
\left\vert \phi \right\vert^{6}$-BIHM models introduced in this work is to
made the vortex even more thicker (green and orange curves, corresponding to
models (\ref{eq:model2}) and (\ref{eq:model3}), respectively). This is also
reflected in the physical magnitudes characterizing the vortex, as both the
magnetic field and the energy density profiles (see Figs. \ref{magnetic_p}
and \ref{energy_p}, respectively) undergo large modifications as compared to
their standard counterparts. In general this means that, at fixed $\beta$,
one can control thickness and physical magnitudes of the vortex by
introduction of suitable $G$ and $\omega$ functions.


Hereafter, we depict the profiles for the second and third models by fixing $%
n=1$ and some values of $\beta$. From Figs. \ref{higgs_2_3} and \ref%
{vetorial_2_3} we see that for $\left\vert \phi \right\vert^{4}$-BIMH model
the thickness of the vortex increases as $\beta$ decreases, i.e., when the
nonlinear effects of the Born-Infeld contribution grow stronger, while for
the $\left\vert \phi \right\vert^{6}$-BIMH model the new effects play a very
little role, leaving almost unmodified the vortex profile. For the $%
\left\vert \phi \right\vert^{4}$-BIMH model this implies large modifications
on the magnetic field and energy density profiles, since their maximum at $%
r=0$ grows quickly with $1/\beta$. On the other hand, as one could have
expected, the tiny modifications on the vortex profile with $\beta$ in the $%
\left\vert \phi \right\vert^{6}$-BIMH model also have little effect on the
magnetic field and energy density profiles. For this model these profiles
have a different behavior as in the $\left\vert \phi \right\vert^{4}$-BIMH
model, since their maxima are not attained at $r=0$, but rather at a finite
distance, a feature that holds for any value of $\beta$.

This analysis shows that the modified-BIHM models through $\beta$
corrections do not change the qualitative features of the physical
magnitudes characterizing the vortex, but are able to introduce quantitative
modifications, which can become large, as in the $\left\vert \phi
\right\vert^{4}$-BIMH model.

\section{Conclusions}

In this work we have studied a family of generalized Born-Infeld theories
with a free parameter, $\beta $, and three generalizing functions which are
nonnegative. These generalizing functions are constrained by the condition (%
\ref{xcxc}) which is the Ampere law of the model. We have worked out the
theory and obtained BPS solutions of vortex-type using Bogomol'nyi trick and
determined the physical properties of the solutions in terms of the magnetic
flux and energy density. It was shown that whenever the conditions (\ref{eq:constraint}) are satisfied, the energy of the topological
vortices has a lower bound
\begin{equation}
E\geq 2\pi v^{2}\left\vert n\right\vert ,
\end{equation}%
which is saturated by the self-dual or BPS topological solutions. In the
numerical analysis we have employed two classes of models characterized by
the potential term, namely, $|\phi |^{4}$ and $|\phi |^{6}$ models, and
depicted the corresponding results for the field profiles and the physical
magnitudes characterizing the vortices. Such results have been compared to
those of the standard Maxwell-Higgs, Chern-Simons-Higgs and
Born-Infeld-Maxwell-Higgs models.

As observed in other cases of Born-Infeld-type modifications in the
literature, the introduction of finite values for the Born-Infeld $\beta $
has a non-trivial impact on the field profiles of the vortices, with the
result that the corresponding physical properties can be controlled by
adequate combination of Born-Infeld modification and $\omega (g)$ and $G(g)$
functions. When we vary $\beta $, however, the size of the variation of the
vortex properties largely depends on the model chosen, with the $|\phi |^{4}$
one showing important variation, while the $|\phi |^{6}$-one is almost
insensitive to changes in $\beta $. Since topological defects find
applications to many context of modern physics as useful tools for the
modelling of different kinds of systems, to be able to modify the physical
properties of vortex solutions is a strong motivation in favour of
consideration of this kind of models. Finally let us mention that the
parameter $\beta $ can not be made arbitrarily small. Our numerical analysis
shows that for all $|\phi|^4$-models the solutions are obtained when $%
\beta>1 $. In the case of the $|\phi|^6$-models it was observed that when $%
\beta\geq 1$ the numeric computations are always valid. The presence of a
critical minimum value, $\beta_c$, below which numerical computations break
down and no solution can be attained, seems to be a quite general phenomenon
occurring in Born-Infeld type modifications, as found in other
investigations in the literature \cite{Hora_Rubiera,Moreno, Bri}. In those
cases, around $\beta_c$ the physical magnitudes characterizing the
topological defect change abruptly as $\beta$ is slowly varied, as happens
in our case. Though some research has been performed about the implications
of this feature, this issue remains unsolved. To conclude, we point out that
the results presented here could be generalized to include non-symmetric BPS
fields.

\section*{Acknowledgments}

R.C. thanks CNPq, CAPES and FAPEMA (Brazilian agencies) by financial
support. D.R.-G. is supported by the NSFC (Chinese agency) grants No.
11305038 and 11450110403, the Shanghai Municipal Education Commission grant
for Innovative Programs No. 14ZZ001, the Thousand Young Talents Program, and
Fudan University, and acknowledges partial support from CNPq grant No.
301137/2014-5.

\end{document}